  \documentclass[11pt]{article}

  \usepackage{latexsym}
  \usepackage{amssymb}
  \usepackage{graphicx}
 
  \def\AFOUR{%
  \setlength{\textheight}{9.0in}%
  \setlength{\textwidth}{5.75in}%x

  \setlength{\topmargin}{-0.375in}%-0.375in
  \hoffset=-0.5in % -.5in %
  \renewcommand{\baselinestretch}{1.17}%
  \setlength{\parskip}{6pt plus 2pt}%
  }

  \AFOUR                                          %%%    ***

  \makeatletter
  \def\section{\@startsection {section}{1}{\z@}
  {-3.5ex plus -1ex minus   -.2ex}{2.3ex plus .2ex}{\large\bf}}
  \def\subsection{\@startsection{subsection}{2}{\z@}
  {-3.25ex plus -1ex minus  -.2ex}{1.5ex plus .2ex}{\normalsize\bf}}
   \def\subsubsection{\@startsection{subsubsection}{3}{\z@}
  {-3.ex plus -1ex minus -.2ex}{1.ex plus .2ex}{\normalsize\it}}
   \makeatother

  \makeatletter
  \@addtoreset{equation}{section}
  
  \makeatother

 \newtheorem{theorem}{Theorem}[section]
\newtheorem{lemma}{Lemma}[section]

\begin{document}

  \global\parskip=4pt

 %%%%%%%%% title page %%%%%%%%%%%%%%%%%%%%%%%%%%

 \makeatletter%
 \begin{titlepage}
         \begin{center}
 \vskip .5in
 \begin{flushright}  ICMPA-MPA/2008/15 \\
LPT-Orsay 08-50
 \end{flushright}

 \begin{center}
 \vskip .5in

 {\LARGE\bf  Color Grosse-Wulkenhaar models: 
\\ One-loop $\beta$-functions\footnote{Work supported
by the ANR Program ``GENOPHY" and by the  Daniel Iagolnitzer Foundation, France. } }
 \end{center}

  \begin{center}
 {\bf Joseph Ben Geloun$^{a,b}$ and Vincent Rivasseau$^{c}$ }\\

$^{a}$International Chair of Mathematical Physics and Applications\\ 
ICMPA-UNESCO Chair, Universit\'e d'Abomey-Calavi,
Cotonou, Rep. of Benin\\
$^{b}$Facult\'e des Sciences et Techniques \\
Universit\'e Cheikh Anta Diop, Dakar, S\'en\'egal\\
Email: joseph.bengeloun@cipma.uac.bj, \  jobengeloun@gmail.com\\
$^{c}$Laboratoire de Physique Th\'eorique, UMR CNRS 8627\\
Universit\'e Paris-Sud X1, 91405 Orsay, France\\
Email: Vincent.Rivasseau@th.u-psud.fr, \ vincent.rivasseau@gmail.com
 \vspace{0.5cm}

    \end{center}

\vspace{10pt}
 \today
 \begin{abstract}
The $\beta$-functions of $O(N)$ and $U(N)$ invariant Grosse-Wulkenhaar models are computed at one loop
using the matrix basis. In particular, for ``parallel interactions", the model is proved asymptotically free in 
the UV limit for $N >1$, and has a triviality problem or Landau ghost for $N<1$. The vanishing $\beta$-function 
is recovered solely at $N=1$.  We discuss various possible consequences of these results.
 \end{abstract}

\noindent
\end{center}
Pacs numbers: 02.40.Gh, 11.10.Nx.

Key-words:  Noncommutative field theory, Grosse-Wulkenhaar model, renormalization, 
effective expansion, $\beta$-function.
 \end{titlepage}
 \vspace{500pt}
 \makeatother

%%%%%%%%% end of title page %%%%%%%%%%%%%%%%%%%%%%%%%%%%%%

\section{Introduction}
\label{Sect1}

Noncommutative (NC) quantum field theory (NCQFT) \cite{Doug} receives an increasing attention
of the theorist community since the advent of a class of  renormalizable theories
built around the Grosse and Wulkenhaar model (GWm) \cite{GW,Riv4}. 
This model is a  NC $\phi^4_4$ scalar field theory on the Moyal-Weyl Euclidean space
with a particular modification of the propagator. 
The GWm is dual in the sense of Langmann-Szabo (LS) \cite{LS},
i.e. considers in a dual manner positions and momenta. This requires the inclusion of an harmonic term 
in  the ``naive" NC $\phi^4$ theory which allows  to circumvent  the deadlock of UV/IR mixing
and to insure the renormalizability at all orders of perturbations \cite{GW,Riv}.

As a corollary, a series of fascinating facts about the GWm have been highlighted  \cite{Riv4}, \cite{Riv}-\cite{diser2}
using different field theory techniques, and other models have been proved renormalizable
using the same ideas \cite{Vign}. More specifically, the study of the 
renormalization group (RG) flow has been investigated in detail \cite{diser}\cite{diser2}.
The subsumption of a vanishing $\beta$-function, obtained at one loop in \cite{GW2},  has been finally
achieved at all orders in \cite{diser2} (with $\Omega=1$) with as major consequence
that the GWm has no Landau ghost (Lg) or triviality problem. The Lg means that, if the renormalized
coupling constant is kept fixed and small, the bare constant increases without apparent bound
as the UV cutoff is removed.
Equivalently, triviality means that the renormalized coupling constant vanishes if 
the bare constant is kept bounded and small as the UV cutoff is removed. 
These twin diseases affect all quantum field theories except non Abelian gauge theories, 
and it was completely unexpected that they do not occur in the GWm.  
Moreover, the theory is not asymptotically free either, but asymptotically safe: 
both bare and renormalized coupling constants remain bounded 
after removing  the UV cutoff.  A nice argument in order to understand the absence 
of the usual  ``charge screening" phenomenon is the following \cite{diser}.  At one loop, 
the wave function renormalization (wfr) does not vanish in contradistinction with the wfr of the commutative 
theory\footnote{Indeed, the commutative tadpole is local and then 
induces a null contribution to the wfr.}.
Taking into account this wfr,  the $\beta$-function \emph{hitherto }positive for $\Omega<1$,  
tends to zero as $\Omega$ tends to $1$. In a way, the RG flow grinds to a halt at $\Omega=1$
because the LS-duality at that point renders perfectly indistinguishable positions and momenta.
The above scenario called ``death of Landau ghost"  has been shown to hold to all orders
of perturbations by a combination of Ward identities and Dyson-Schwinger equations \cite{diser2}.
Thus, quantum field theories on NC geometry are, at least in this sense, better behaved
than ordinary ones and the GWm is a promising candidate for the constructive program 
\cite{Riv7}.

$O(N)$ and $U(N)$  invariant NC models with a $N$-valued color index have 
been considered in \cite{ncol}, and the vacuum and symmetry breaking of GW models of this type 
has been investigated in \cite{dego}.
In this paper, we investigate the $\beta$-functions of  such models.
It should be emphasized that the limits $N\to 0$ and $N\to \infty$ 
could be both of special interest,
for they are related, respectively, to polymers with non-local self-avoiding interactions \cite{Genn}
and to a kind of solvable spherical  \cite{Baxt} GWm.
We focus, hereunder, on the one-loop computations
recalling  that the behavior of the coupling constant flow 
is really determined by the sign of the first non vanishing coefficient of the $\beta$-function.
In contrast, with $N=1$, we find non zero coefficients at one loop order, so
without further computations, we can reach conclusions on the UV behavior of these
rotation invariant GW models. For the class of $O(N)$  models with ``parallel interaction",
we find that such models are asymptotically free in the UV regime for $N>1$,
and hence also susceptible of a full constructive analysis.
The model at $N=0$ is not asymptotically free.
The $U(N)$ invariant complex NC $\phi^4_4$ theory is also 
discussed and has similar features according to the value of $N$.
Finally let us mention that asymptotically safe models of the GWm type with 
vanishing $\beta$-function at all orders can be obtained by adding a magnetic
field \cite{BGR}.

The paper is organized as follows. The next section is devoted
to notations and general considerations for the $O(N)$ GWm.
The complex case construction and treatment are deferred in Section 3 
as a direct generalization of the real analysis.
We assume the knowledge of renormalization and 
effective expansions as developed for instance in \cite{Riv7},
and of NC field theories in the matrix basis as treated in \cite{GW,diser,diser2}.
Section 3 gives our main result, its  proof and further discussions.
The conclusion of this study is given in  Section 4.

\section{The $O(N)$ Grosse-Wulkenhaar model: Notations and considerations}
\label{Sect2}

Let us consider a real vector field $(\phi^a)_{a=1,\dots, N}$ theory,  with $N\in \mathbb{N}$,
defined by the GWm action in the NC Euclidean spacetime $\mathbb{R}^{4}$ \cite{GW}
\begin{eqnarray}
S&=& \int d^4x\,\,\left\{ \sum_{a=1}^{N} \left( 
\frac{1}{2}\partial_{\mu}\phi^a \star \partial^{\mu}\phi^a +\frac{\mu}{2}(\phi^a)^2
+ \frac{\Omega}{2} (\tilde{x}\phi^a)^2 \right) \right. \cr
&& \left.  + \frac{\lambda_1}{2} \sum_{c,d=1}^{N}\phi^{c}\star \phi^{c}\star\phi^{d}\star\phi^{d}
+ \frac{\lambda_2}{4}\sum_{c,d=1}^{N}\phi^{c}\star \phi^{d}\star\phi^{c}\star\phi^{d}\right\} ,
\end{eqnarray}
where $\tilde{x}_\nu = 2(\theta_{\nu\mu}^{-1})x^{\mu}$, $\theta_{\nu\mu}^{-1}$ being the inverse
of the anticommutative matrix associated with the Moyal $\star$-product.
Two coupling constants $\lambda_1$ and $\lambda_2$  have been introduced 
for the two natural interactions in the quartic term. 
Now, we write the theory in the matrix basis and use the simpler normalizations of 
Ref.\cite{diser2}, with $\Omega=1$\footnote{For any $N\ge 0$, we can check in (\ref{omegaflow})
that indeed the RG flow leads to $\Omega = 1$ in the UV regime as in the $N=1$ case.}.
The bare propagator is given by, for all  $m,\,n,\,k,\,l\, \in \mathbb{N}^2$ 
and $ \delta_{mn} := \delta_{m_1n_1} \delta_{m_2n_2}$,
\begin{eqnarray}
C_{mn;kl}= G_{mn}\, \delta_{ml}\delta_{nk};\quad
G_{mn}=(m+n+A)^{-1}; \quad m+n:=m_1+m_2 + n_1+n_2,
\end{eqnarray}
with $A=2+\mu^2/4$,  and the vertices take the form
(henceforth, implicit sum from $1$ to $N$ over repeated color indices is used)
\begin{equation}
V_1=
\frac{\lambda_1}{2} \sum_{m,n,k,l\in \mathbb{N}^2}\phi_{mn}^{c}\,\phi_{nk}^{c}\,\phi_{kl}^{d}\,\phi_{lm}^{d},
\quad V_2= 
\frac{\lambda_2}{4} \sum_{m,n,k,l\in \mathbb{N}^2} \phi_{mn}^{c}\,\phi_{nk}^{d}\,\phi_{kl}^{c}\,\phi_{lm}^{d}.
\end{equation}
We call $V_1$ the ``parallel" vertex and 
$V_2$ the ``crossed" vertex. 
After these manipulations, the action can be written as
\begin{eqnarray}
S=
\frac{1}{2} \sum_{m,n\in \mathbb{N}^2}  \phi_{mn}^{a}G^{-1}_{mn}\,\phi_{nm}^{a}+
\,\,V_1\,+\,V_2.
\label{act}
\end{eqnarray}

{\bf Renormalizability of the model.}
At $\Omega = 1$, the propagator is diagonal in the color index and independent of 
its value and it is the same as in the ordinary GWm\footnote{For $\Omega = 1$, the model
is a pure matrix theory, in the sense that the probability distribution
is independent for each matrix coefficient  although it is not identically distributed (often called i.ni.d.). 
As already remarked, for any fixed $N \ge 0$,
the flow of $\Omega$ always goes rapidly to $\Omega=1$
in the UV which is of course the small distance limit of interest \cite{diser}.}.
Therefore, a slice decomposition of the propagator identical to the one
of \cite{Riv} leads to an identical power counting.  One concludes that, as in the $N=1$ case,
the only divergent contributions come from graphs with two or four external legs with genus zero
and exactly one broken external face. The techniques of subtraction of logarithmically 
divergent graphs and of mass and wave function renormalizations 
used in the GWm can be applied here. Hence, the $O(N)$ GWm as defined
in (\ref{act}) is renormalizable. 

{\bf Goals.}
We want  to compute at one loop the dynamics of the effective constant couplings, 
say
\begin{eqnarray}\label{gamma4}
\lambda_{1,r} = -\frac{\Gamma_{4, ||}(0,0,0,0)}{Z^2}, \quad\quad
\lambda_{2,r} = -\frac{\Gamma_{4, \times}(0,0,0,0)}{Z^2}, 
\end{eqnarray}
where the wave function renormalization is
\begin{eqnarray}
Z= 1- \left.  \partial_{m_1} \Sigma (m,n) \right|_{m=0=n}
\label{zee}
\end{eqnarray}
and the self-energy $\Sigma(m,n)$ is the sum of the amputated 
one particle irreducible (1PI) amplitudes of the two point correlation function
\begin{eqnarray}
 \Sigma (m,n) = \langle \, \phi_{mn}^{a}\phi_{nm}^{a}  \, \rangle^{t}_{1PI}.
\label{topt}
\end{eqnarray}
The amputated 1PI four point functions in (\ref{gamma4}) are 
\begin{eqnarray}
\Gamma_{4, ||}(m,n,k,l) = 
\langle \, \phi_{mn}^{a}\phi_{nk}^{a} \phi_{kl}^{b}\phi_{lm}^{b}  \, \rangle^{t}_{1PI},
\quad
\Gamma_{4, \times}(m,n,k,l) = 
\langle \, \phi_{mn}^{a}\phi_{nk}^{b} \phi_{kl}^{a}\phi_{lm}^{b}  \, \rangle^{t}_{1PI}.
\label{fopt}
\end{eqnarray}
Note that $a$ and $b$  are fixed in equations (\ref{topt}) and (\ref{fopt}).
Furthermore, the derivative taken on $m_1$ (\ref{zee}) is actually a matter
of choice since one obtains the same result by deriving by the external
indices $m_2, n_1, n_2$ and putting the remaining to $0$.

The flow of the $\Omega$ parameter, at one loop and up to a constant,
is given as in \cite{GW} by
\begin{eqnarray}\label{omegaflow}
\Omega^2_r = \frac{(1-\Omega^2)}{\Omega}
(1- \left.  \partial_{m_1} \Sigma (m,n) \right|_{m=0=n})
\end{eqnarray}
and induces, for any $N \ge 0$,
the UV fixed point $\Omega_{bare} = 1$ (see below Eq.(\ref{sigz})).

{\bf Feynman rules.}
In the scalar GWm, the Feynman rules are expressed in terms of ribbon graphs.
Only graphs with genus $g=0$ and one broken external face 
$B=1$ with at most 4 external legs may diverge. They govern the RG flow of
the parameters $\lambda$, $\mu$, $\Omega$ and the field strength $Z$ (or wfr)
of the model.  Considering  the $O(N)$ model, it is convenient to add to the ribbon graphs an inner 
``thread" or ``decoration" which represents the color index.
The corresponding Feynman rules read off as

\begin{figure}
\centering
    \begin{minipage}[t]{0.45\textwidth}
      \centering
\includegraphics[width=7cm, height=3.5cm]{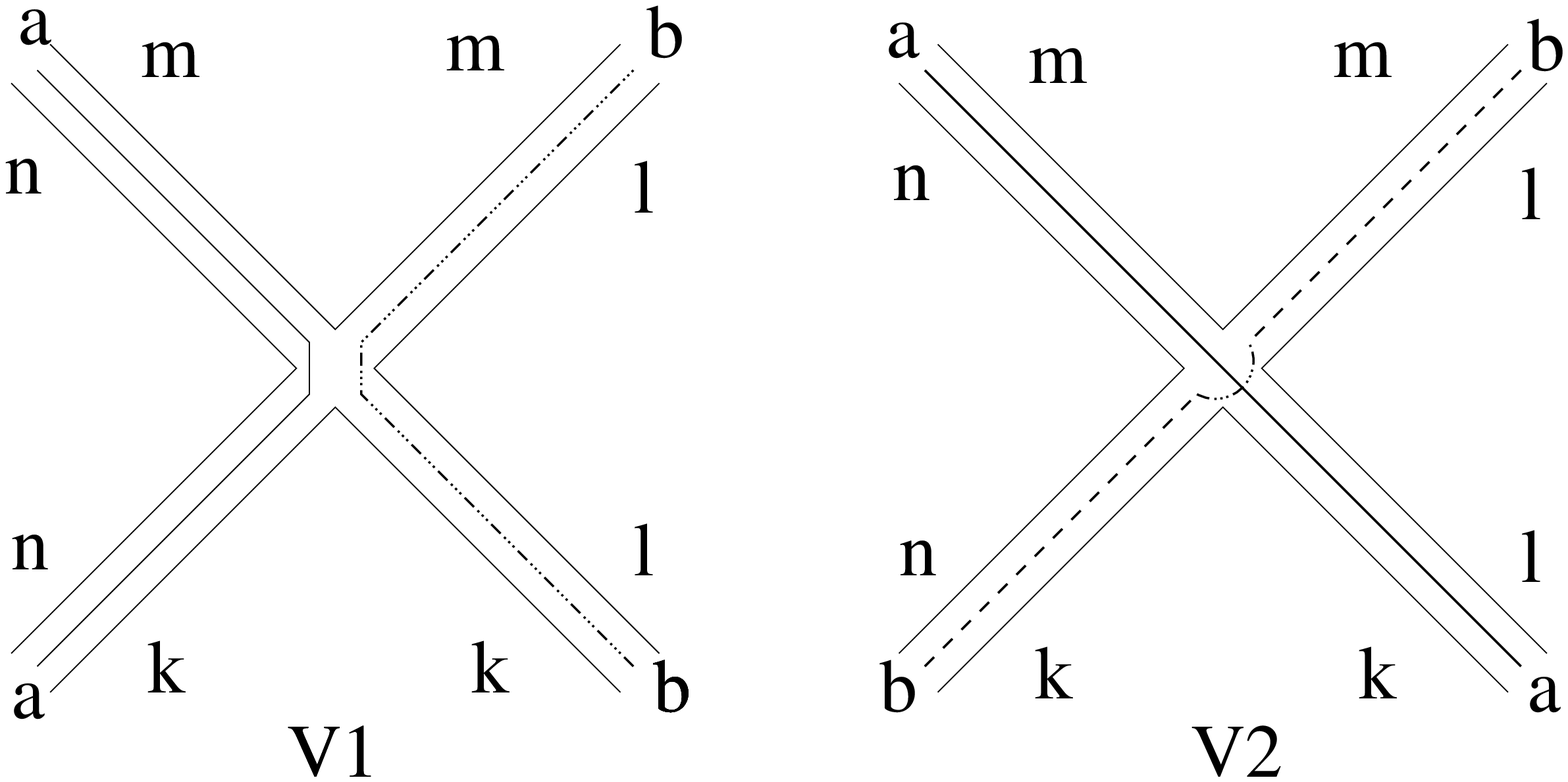}
\caption{ {\small Vertices of type $V_1$ (parallel) and $V_2$ (crossed) of the color model.}}
\label{fig1}
\end{minipage}
\hspace{0.5cm}
\begin{minipage}[t]{0.45\textwidth}
     \centering
\includegraphics[width=7cm, height=4cm]{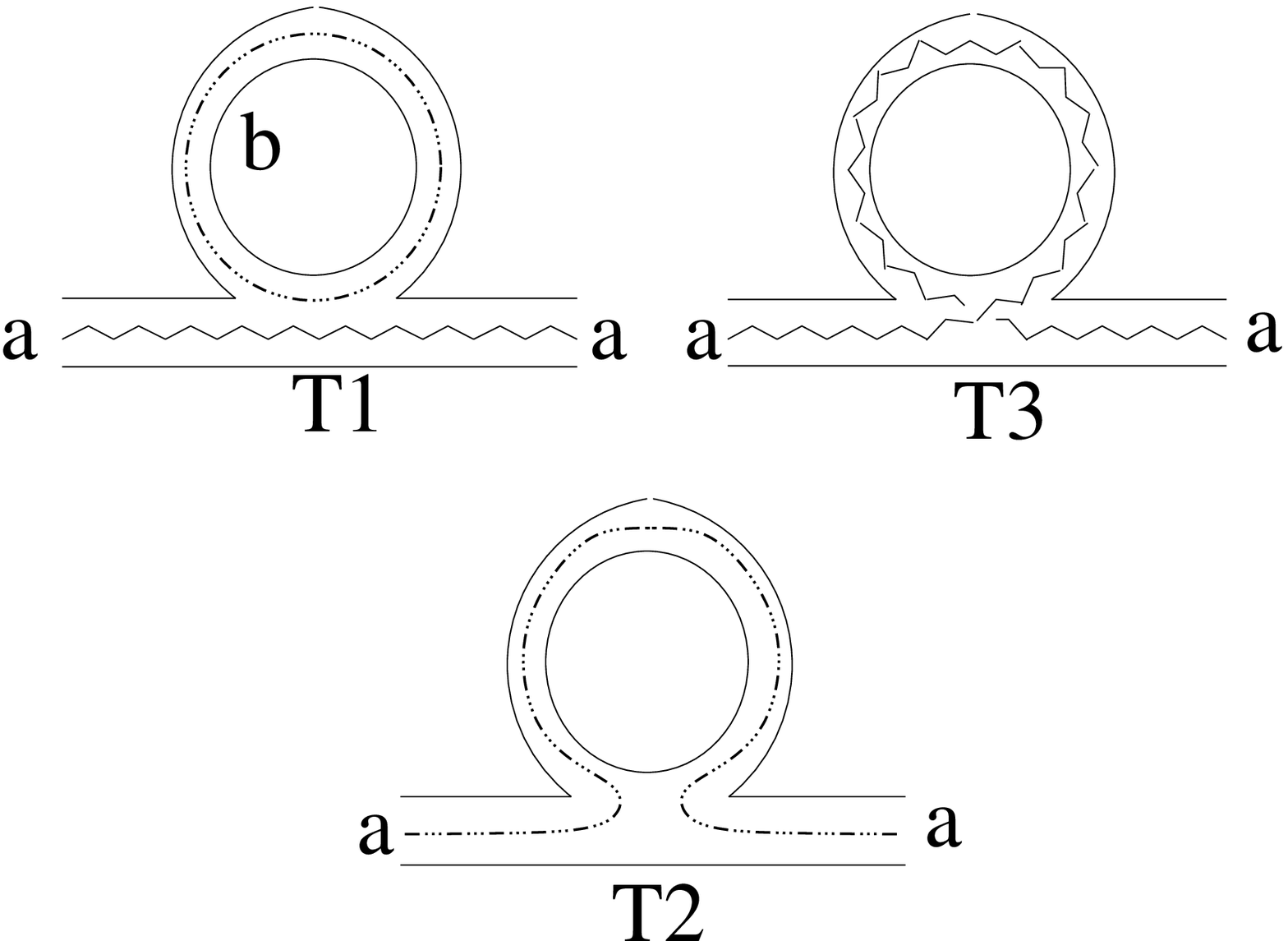}
\caption{{\small Tadpoles $T_1$, $T_2$ and $T_3$ of the model.}}
\label{fig2}
\end{minipage}
\end{figure}

\begin{enumerate}
\item[(i)] To each ribbon line is associated a $G_{mn}$ propagator;
\item[(ii)] Each vertex of the first kind $V_1$ (see Figure \ref{fig1}) 
has a weight  equals to $\lambda_{1}/2$;
\item[(iii)] Each vertex of the second kind $V_2$ (see Figure \ref{fig1}) 
has a weight  equals to $\lambda_{2}/4$.
\item[(iv)] To each ribbon face is associated a sum over the $(m_1,m_2)$ corresponding integers.
These sums together with the $G_{mn}$ propagators command to the usual power counting,
which is the same as the one of the scalar model.
\item[(v)] The color sum produces an additional contribution of $N$ for each colored ``bubble".
For instance, only $T_1$ (see Figure \ref{fig2}) will have a factor of $N$. 
\end{enumerate}

\section{One loop $\beta$-functions}
\label{Sect3}
As was thoroughly argued in \cite{diser} and shall not be refrained here, 
the effective perturbative expansion is better in order to study the $\beta$-function for 
it just contains the necessary subtractions. We therefore use the effective expansions
in the next developments. 
One notes that the sum involved in ensuing amplitude of graphs are divergent. 
It is only after subtraction of the mass divergences and other subtractions 
that these sums appear now finite even after removing the cutoff. 
The following statement holds.
\begin{theorem}\label{theo}
At $\Omega=1$, we have
\begin{eqnarray}
&&
\lambda_{1,r} =
\lambda_{1} - \left( \lambda_1^2(1-N) + \lambda_2^2   \right) S^{(1)}
+ O(\lambda_{i=1,2}^2) + O(\lambda_{1}\lambda_{2}), \label{lam1}\\
&&\lambda_{2,r} 
= \lambda_{2} +\left( 2 \lambda_2^2 - 2(1-N)\lambda_1\lambda_2 \right)S^{(1)}
+ O(\lambda_{i=1,2}^2) + O(\lambda_{1}\lambda_{2}), \label{lam2}
\end{eqnarray}
where $S^{(1)} := \sum_{p\in \mathbb{N}^2} 1/(p+A)^2$ which is logarithmically
divergent when removing the UV cutoff.
\end{theorem}
The divergence of $S^{(1)}$ is logarithmic and corresponds both to the bubble 
four point function divergence and to the 
wave function renormalization of the tadpole {\it after mass subtraction}.
The proof of this theorem involves the coming lemmas.
\begin{lemma}\label{lem1}
At $\Omega=1$, for the NC $(\phi^{a})^4_4$ model at the first order in $\lambda_{i}$, $i=1,2$,
the self-energy and the wave function normalization are
\begin{eqnarray}
&&\Sigma(m,n) = -\left(\lambda_{1}(N+1)+ \lambda_2\right)
\sum_{r\in \mathbb{N}^2} \left( G_{mr} + G_{r n} \right),
\label{eqsig}\\
&&Z= 1- \left. \partial_{m_1}\Sigma(m,n)\right|_{m=0=n} =
1- \left( \lambda_1(N+1)  + \lambda_2 \right)S^{(1)},
\label{sigz}
\end{eqnarray}
respectively.
\end{lemma}
\begin{lemma}\label{lem2}
At $\Omega=1$, for the NC $(\phi^{a})^4_4$ model at one loop,
the amputated 1PI four point functions are given by,
\begin{eqnarray}
&&\Gamma_{4,||}(0,0,0,0) = - \lambda_1 + \left(\lambda_1^2 (N+3) 
+ 2 \,\lambda_1\lambda_2 +\lambda_2^2\right) S^{(1)},\label{gamori1}\\
&&\Gamma_{4,\times}(0,0,0,0) = - \lambda_2 + 4\,  \lambda_1\lambda_2\,  S^{(1)}.
\label{gamori2}
\end{eqnarray}
\end{lemma}
The remaining of this section is devoted to the proof of these
lemmas and theorem.

{\bf Proof of Lemma \ref{lem1}.}
The self-energy is
\begin{equation}
\Sigma(m,n) =\sum_{{\cal G}_i}  K_{{\cal G}_i}S_{{\cal G}_i}(m,n) 
\label{sigma}
\end{equation}
where ${\cal G}_i$ runs over two point 1PI graphs,
$S_{{\cal G}_i}(m,n)$ is its amplitude 
and $K_{{\cal G}_j}$ is the corresponding combinatorial weight including color summation, 
that is the number of Wick contractions given rise to ${\cal G}_j $ times the sum over color indices. 
At first order, only the tadpole graphs $T_1$, $T_2$ and $T_3$ (see Figure \ref{fig2}) contribute to 
(\ref{sigma}) with the combinatorial factors
\begin{equation}
K_{T_1} = 2N,\quad\quad K_{T_2}= 2,\quad\quad
K_{T_3}= 4,
\label{com}
\end{equation}
respectively. We introduce $S^{(1)}(m,n)= \sum_{r\in \mathbb{N}^2} \left( G_{mr} + G_{r n} \right)$ 
and get the amplitudes
\begin{equation}
S_{T_1}(m,n) =  -\frac{\lambda_1}{2} S^{(1)}(m,n) ,\;
S_{T_2}(m,n)=  -\frac{\lambda_1}{2}  S^{(1)}(m,n),\;
S_{T_3}(m,n)=  -\frac{\lambda_2}{4} S^{(1)}(m,n) 
\label{st}
\end{equation}
where the propagators $G_{mr} $ and $ G_{r n} $ coincide with  the up 
and down versions of the tadpole, respectively \cite{diser}.
Equation  (\ref{eqsig}) follows from (\ref{com}) and (\ref{st}).
The wfr $Z$ is readily obtained by taking the
correct derivative onto $\Sigma(m,n)$ and setting external indices to zero.
\hfill $\square$

\begin{figure}
\centering
    \begin{minipage}[t]{0.8\textwidth}
\includegraphics[width=12cm, height=5cm]{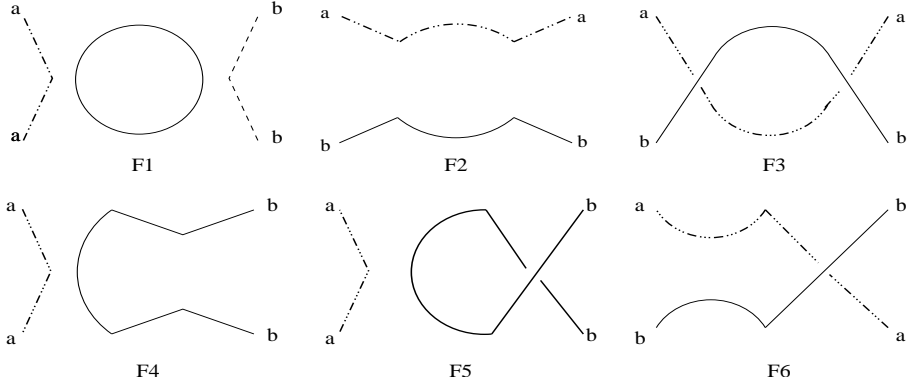}
\caption{ {\small Run of colors: One loop 1PI decorated graphs $\{F1,F2,F3,F4,F5\}$ and $\{F_6\}$ 
associated with four point functions  $\Gamma_{4,||}$ and  $\Gamma_{4,\times}$, respectively.}}
\label{fig3}
\end{minipage}
\end{figure}

{\bf Proof of Lemma \ref{lem2}.}
In order to determine the two four point functions driving the equations
of the RG flows at one loop, we consider the ``decorated" graphs 
of Figure \ref{fig3}. Note that, only the run of colors in the vertices
is represented. Given the ``skeleton"  (this stands for the set of 
cyclic ordering of the external indices $(m,n,k,l)$ connected on
the ribbon one loop diagram \cite{diser} in a planar manner without more than one 
external broken face and a genus zero as it is the case in the scalar GWm), 
the data really sufficient for the underlying combinatorics of the perturbative study 
is the above color course. The one loop 1PI amputated four point functions have the form
\begin{eqnarray}
\Gamma_{4,(\cdot)}(m,n,k,l)=\sum_{{\cal G}_i}  K_{{\cal G}_i}{\cal S}_{{\cal G}_i}(m,n,k,l),
\end{eqnarray}
here,  ${\cal G}_i$  are four point 1PI graphs with the required topology
of  amplitude ${\cal S}_{{\cal G}_i}(m,n,k,l)$ 
and of combinatorial weights $K_{{\cal G}_j}$. 
The graphs $F_{i}$, $i=1,2,3,4,5$, contribute to $\Gamma_{4,||}$
and $F_6$ to $\Gamma_{4,\times}$ (Figure \ref{fig3}) 
and their combinatorial factors (including again color summation) are
\begin{equation}
K_{F_1} =2^3 \cdot  N, \;\; K_{F_2}=2^3, \;\; K_{F_3}= 4^2 \cdot 2,\;\; K_{F_4}=2^3\cdot 2,\;\;
K_{F_5}=(4 \cdot 2)\cdot 2,\;\;K_{F_6}=(4 \cdot 2)\cdot 2.
\label{com4}
\end{equation}
It is worthy to emphasize that each graph $F_i$, for $i=4,5$, has a symmetric partner
with the same amplitude, so that their combinatorial factor $K_{F_i}$, $i=4,5$, have taken 
into account such a symmetry factor. Things appear differently for the graph $F_6$ 
which has four partners by rotating the graph by $\pi/2$ but the amplitudes generated
by these rotations are not all equal. 
The combinatorial factor $K_{F_6}$ considers only the rotation by $\pi$.
We get the amplitudes such that
\begin{eqnarray}
&{\cal S}_{F_1}(m,n) = \tilde\lambda_{1}\,{\cal S}(m,k),\;\;
{\cal S}_{F_2}(n,l)=  \tilde\lambda_{1}\,{\cal S}(n,l),\;\;
{\cal S}_{F_4}(m,n) =    \tilde\lambda_{1}\, {\cal S}(m,k),\label{s1}\\
&{\cal S}_{F_3}(m,k)= \tilde\lambda_{2}\,{\cal S}(m,k),\quad
{\cal S}_{F_5}(m,k) = \tilde\lambda_{12} \,{\cal S}(m,k),\label{s2}\\
&{\cal S}_{F_6}(m,n,k,l) = \tilde\lambda_{12}\,({\cal S}(m,k)+{\cal S}(n,l) ),\label{s3}\\
&\tilde\lambda_{i}:=\lambda^2_i/(2!\, 2^2),\;\;\;i=1,2,\quad
\tilde\lambda_{12}:= \lambda_1\lambda_2/(2\cdot 4), \nonumber\\
& \forall m,k \in \mathbb{N}^2,\quad
{\cal S}(m,k):= \sum_{r\in \mathbb{N}^2} \frac{1}{(m+r+A)}\cdot  \frac{1}{(k+r+A)}.
\nonumber
\end{eqnarray}

Given (\ref{com4}) and (\ref{s1})-(\ref{s3}),
the following expressions rest on a straightforward algebra 
\begin{eqnarray}
\Gamma_{4,||}(m,n,k,l)&=& -\lambda_{1} + \lambda_{1}^2\left( (N+2){\cal S}(m,k) + {\cal S}(n,l) \right)\cr
&+& 2 \lambda_1 \lambda_2\, {\cal S}(m,k) +\lambda^2_{2} \,{\cal S}(m,k),  \label{gam1}\\
\Gamma_{4,\times}(m,n,k,l) &=&-\lambda_2 + 2 \lambda_1 \lambda_2\, \left( {\cal S}(m,k) + {\cal S}(n,l)  \right),
\label{gam2}
\end{eqnarray}
The lemma is proved from (\ref{gam1}) and (\ref{gam2}), once the external indices 
$m,\,n,\,k,\,l$ are put to $0$. \hfill$\square$

{\bf Proof of Theorem \ref{theo}.}
The proof of equations (\ref{lam1}) and (\ref{lam2}) come out from the quotients
\begin{eqnarray}
\lambda_{r,1}&=&-\frac{\Gamma_{4,||}(0,0,0,0)}{Z^2} = 
-\frac{ - \lambda_1 + \left(\lambda_1^2 (N+3) + 2 \,\lambda_1\lambda_2 +\lambda_2^2\right) S^{(1)}}
{\left(1- \left( \lambda_1(N+1)  + \lambda_2 \right)S^{(1)}\right)^2}\cr\cr
&=&   \lambda_1 -  \gamma^{||}_1 \lambda_1^2  -  \gamma^{||}_{12} \lambda_1\lambda_2  -  \gamma^{||}_2 \lambda_2^2  + O(\lambda_{i=1,2}^2 ) +  O(\lambda_1 \lambda_2 ),
\label{run1}\\\cr
\lambda_{r,2}&=&-\frac{\Gamma_{4,\times} (0,0,0,0) }{Z^2} =
-\frac{- \lambda_2 + 4\,  \lambda_1\lambda_2\,  S^{(1)}}{\left(1- \left( \lambda_1(N+1)  + \lambda_2 \right)S^{(1)}\right)^2}\cr\cr
&=&   \lambda_2 -  \gamma^{\times}_1 \lambda_1^2  -  \gamma^{\times}_{12} \lambda_1\lambda_2 -  \gamma^{\times}_2 \lambda_2^2  + O(\lambda_{i=1,2}^2 ) + O(\lambda_1 \lambda_2 ).
\label{run2}
\end{eqnarray}
Expanding the rational function $1/Z^2$ up to the second order in the coupling constants,
we get the coefficients $\gamma^{||,\times}_k$ as in (\ref{lam1}) and (\ref{lam2}), i.e.
\begin{eqnarray}
&&\gamma^{||}_1 =(1-N) ,\;\;\;   \gamma^{||}_{12}=0  ,\;\;\;  \gamma^{||}_2=1, \\
&&\gamma^{\times}_1= 0,\;\;\;   \gamma^{\times}_{12}= 2(1-N) ,\;\;\;  \gamma^{\times}_2=-2 .
\end{eqnarray}
We are then lead to the one loop $\beta$-coefficients
\begin{equation}
\beta_{||}=(1-N) ,\quad\quad \beta_{\times}=-2,
\end{equation}
which achieve the proof of the statement. \hfill$\square$.

{\bf Discussion.}
For $N>1$, one notes that $\beta_{||}<0$. 
Consequently,  in the UV limit and setting $\lambda_{2}=0$,
the model is asymptotically free, i.e. the bare coupling $\lambda_{1}$ is 
screened in this limit. The renormalon problem has to be considered but its contribution possesses
an alternate sign which in principle could be handled by some summability principle
such as Borel resummation \cite{Riv7}.
For $N<1$, and still $\lambda_{2}=0$,  the ordinary issue of Lg arises.
Besides, if we assume that second order terms in equations (\ref{run1}) and (\ref{run2}) vanish,
a way to force the system to be such that $\lambda_{r,i}=\lambda_{i}$, 
the algebraic system reached can be inverted with nontrivial (not free theory) solutions
\begin{equation}
[N=1,\;\;\lambda_{2}=0],\quad\quad[N=2;\;\; \lambda_{1}=- \lambda_{2}].
\end{equation}
The left hand side parameters can be considered as the fixed point of the GWm whereas,
albeit unstable due to one of its sector, the second set of parameters and couplings 
defines a new renormalizable color model which is asymptotically safe and does not
suffer of Lg. Further, this $N=2$ model is not  equivalent to the complex GWm \cite{diser}
as one can immediately check by expanding the action in 
real and imaginary parts of the complex field $\phi=\phi_1 +i \phi_2$.
As a consequence, although being somewhat a vector theory,
the complex renormalizable GWm cannot be interpreted as a $O(N)$ color model
in the sense that we have defined it here.

{\bf The model limit $N=0$.}
As previously claimed, the model limit $N=0$ can be interpreted as
a model of polymers with nonlocal repulsing interaction.
Using the explicit form of the Moyal kernel in position space \cite{Riv2},
we see that what is suppressed is no longer when the polymer chain crosses itself, as in the
usual commutative case, but when four points in the polymer chain
sit at the corners of a parallelogram, and the suppression factor is really an oscillation
proportional to the area of that parallelogram. This may seem completely unphysical, 
but in dimension $2$, the Moyal 
geometry is really the one induced by a constant magnetic field perpendicular
to the plane (see \cite{Riv4} and references therein). 
Renormalizable Moyal interactions may be selected by a RG process,
so that such models may be physically relevant to  the growth
of two dimensional charged polymers under strong
magnetic field. This model can therefore be 
thought as some kind of ``polymer version" of the quantum Hall effect \cite{Riv4}.

Let us rapidly discuss the features of the $\beta$-function of this model at one loop
with respect to the above calculations.
Setting $N=0$ in (\ref{lam1}) and (\ref{lam2}), we have the response
\begin{eqnarray}
&&
\lambda_{1,r} =
\lambda_{1} - \left( \lambda_1^2 + \lambda_2^2   \right) S^{(1)}
+ O(\lambda_{i=1,2}^2) + O(\lambda_{1}\lambda_{2}), \label{lam3}\\
&&\lambda_{2,r} 
= \lambda_{2} +\left( 2 \lambda_2^2 - 2\lambda_1\lambda_2 \right)S^{(1)}
+ O(\lambda_{i=1,2}^2) + O(\lambda_{1}\lambda_{2}). \label{lam4}
\end{eqnarray}
Still, the trivial problem and Lg arises for $\beta_{||}>0$ ($\lambda_2=0$)
but the model is asymptotically free in the infrared direction,
a reminiscent behavior of the ordinary commutative $\phi^4$ theory.

{\bf The model limit $N=\infty$.}
If the index color $N$ tends to infinity independently of the matrix indices, we need to rescale
the coupling $\lambda_i$ into $\lambda_i /N$ to get a non trivial limit.
A further Wick-ordered interaction cancels the tadpoles such that only 
chain of bubble graphs survive and form an explicit computable geometric series.
We recover thence the usual integrability of the so called spherical model.
The $\beta$-function is obtained by taking the $N \to \infty$ in equations (\ref{sigz})
and (\ref{gamori1}) with adapted constant couplings. We find $(\lambda_2=0)$
\begin{equation}
- \lambda^{\infty}_{\;\,\;r,1} = - \lambda_1 -  \lambda_1^2 + O(\lambda_1^2).
\end{equation}
The model then is asymptotically free as expected since we had asymptotic freedom 
for $N >1$.

Far more interesting would be to investigate ``double limits"
in which both the matrix indices $m, n$ and the color index $a$ are sent to infinity
in a coupled way. This may completely change the UV behavior of the theory.
 For instance, we typically no longer have renormalizability in $D=4$ (unless we 
also change the propagator dependence on the color index).
Remark that the crossed vertex $\lambda_2$ is similar to the Barrett-Crane vertex of the 2+1
dimensional group field theory approach to quantum gravity \cite{revgrav}. Thus, models with such 
double limits clearly deserves further study.

\begin{figure}
\centering
    \begin{minipage}[t]{0.6\textwidth}
      \centering
\includegraphics[width=9cm, height=3.5cm]{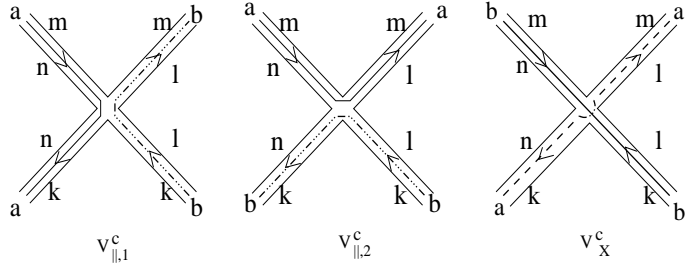}
\caption{ {\small Vertices $V^{c}_{||,1}$, $V^{c}_{||,2}$ 
and $V^{c}_{\times}$ of the complex model.}}
\label{fig4}
\end{minipage} 
\end{figure}

\begin{figure}
\centering
 \begin{minipage}[t]{0.45\textwidth}\centering
\includegraphics[width=6cm, height=3.75cm]{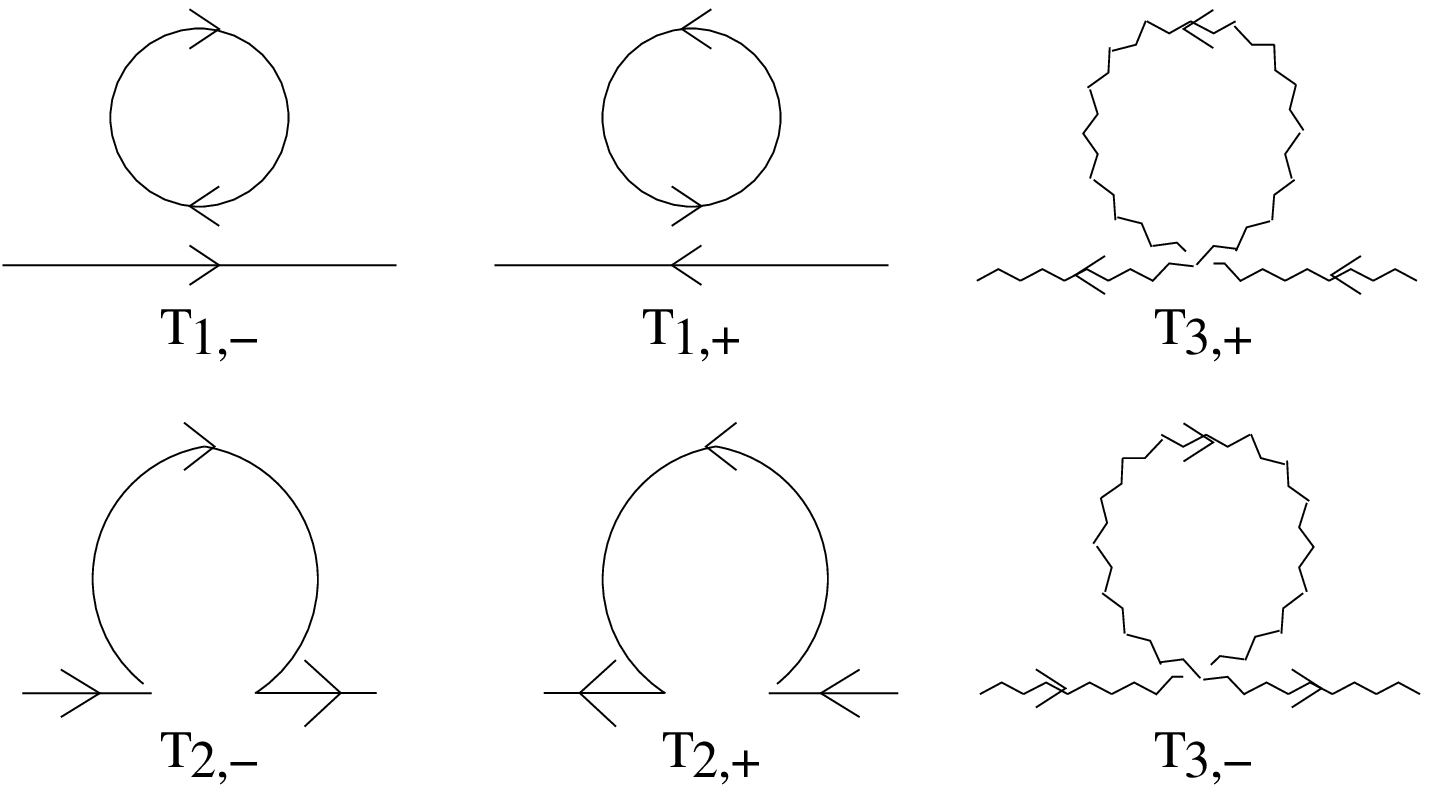}
\caption{ {\small $U(N)$ tadpole diagrams.}}
\label{fig5}
\end{minipage} 
\hspace{0.3cm}\centering
  \begin{minipage}[t]{0.45\textwidth}
\includegraphics[width=6cm, height=2.25cm]{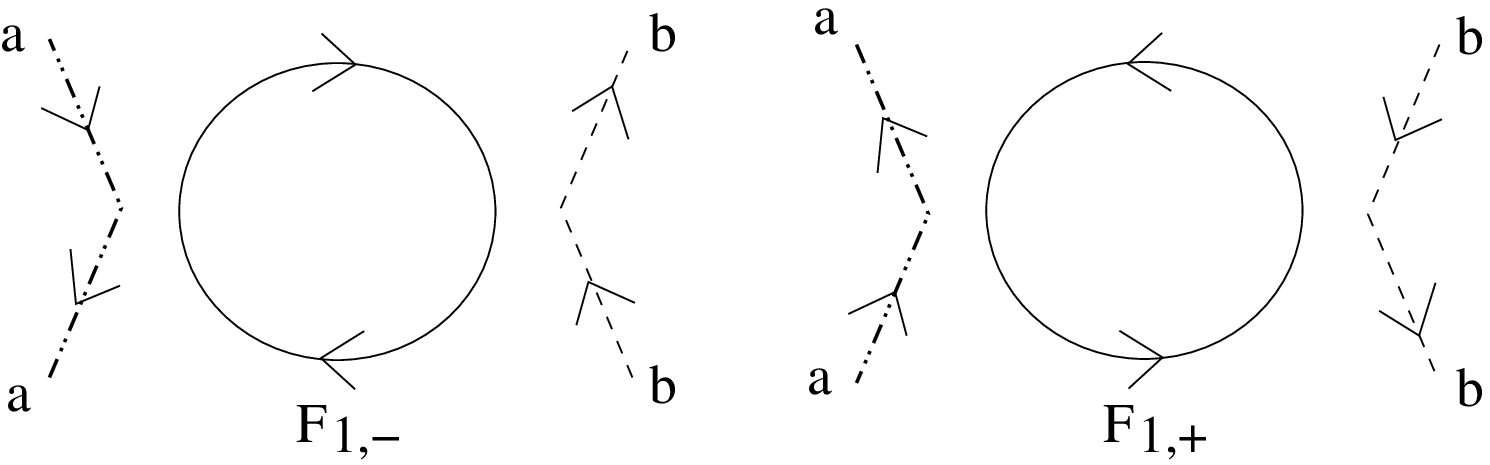}
\caption{ {\small  $U(N)$ "bubble" diagrams.}}
\label{fig6}
\end{minipage}
\end{figure}

{\bf The $U(N)$ NC $\phi^4_4$ theory.}
By an extension
of the previous formulation, we can easily discuss
some features of the $U(N)$ invariant complex NC $\phi^4_4$ model.
Let us recall that the ordinary complex GWm has some peculiarities of cyclic orientation and 
restricted kinds of contractions between $\bar\phi$ and $\phi$ \cite{diser,diser2}.
Given $(\phi^a)_{a=1,...,N}$ a complex vector field with complex conjugated 
$(\bar\phi^{a})^T_{a=1,...,N}$, compiling the complex constraints onto a color model, 
the next vertices are found (see Figure \ref{fig4} for ribbon representations)
\begin{eqnarray}
V^{c}_{||,1}= \frac{\lambda_{1}}{2} \sum_{m,n,k,l\in \mathbb{N}^2}\bar\phi_{mn}^{a}\,\phi_{nk}^{a}\,\bar\phi_{kl}^{b}\,\phi_{lm}^{b},&&
V^{c}_{||,2}= \frac{\lambda_{2}}{2} \sum_{m,n,k,l\in \mathbb{N}^2}\bar\phi_{mn}^{a}
\,\phi_{nk}^{b}\,\bar\phi_{kl}^{b}\,\phi_{lm}^{a},\\
V^{c}_{\times}= \frac{\lambda_{\times}}{2} \sum_{m,n,k,l\in \mathbb{N}^2}\bar\phi_{mn}^{a}\,\phi_{nk}^{b}\,\bar\phi_{kl}^{a}\,\phi_{lm}^{b}.
&&
\end{eqnarray}
The $U(N)$-version  of the action as set in \cite{diser,diser2}
can be written as, with the matrix operator $X_{mn}=m\,\delta_{mn}$,
\begin{eqnarray}
S= \frac{1}{2} \sum_{n,m\in \mathbb{N}^2} (\bar\phi^{a}_{mn} X_{mn} \phi_{nm}^{a} +
\phi^{a}_{mn}  X_{mn} \bar\phi_{nm}^{a} ) +V^{c}_{||,1} +V^{c}_{||,2}+ V^{c}_{\times} .
\end{eqnarray}
The Gaussian measure has the same covariance (propagator) as in the real theory.
Arguments towards  the effective expansions and computation 
of the $\beta$-functions ($\beta_{||,1}$, $\beta_{||,2}$, $\beta_{\times}$) naturally follow.
In a similar way that we treated the $O(N)$ model, 
we compute, on one side, the self-energy $\Sigma^c(m,n)$ associated with
two point amputated 1PI and the wfr $Z^c=1-\partial \Sigma^c(0,0)$
and, on the other side, the four point 1PI amputated amplitude
$\Gamma^c_{4,\sigma,i}$, $\sigma=||,\times$, $i=1,2,\emptyset$,
in order to evaluate at first order the RG flows
\begin{eqnarray}
\lambda^c_{r,\sigma,i}= \frac{\Gamma^c_{4,\sigma,i}(0,0,0,0)}{(Z^c)^2},
\end{eqnarray}
where
\begin{eqnarray}
\Sigma^c(m,n) = \sum_{{\cal G}_i} K^c_{{\cal G}_i} S_{{\cal G}_i}(m,n), \quad
\Gamma^c_{4,\sigma,i}(m,n,k,l) = \sum_{{\cal G}_{\sigma,i}} 
{\cal K}^c_{{\cal G}_{\sigma,i}} {\cal S}_{{\cal G}_{\sigma,i}}(m,n,k,l).
\end{eqnarray}
The ribbon graphs in the complex case are very similar to the real case.
However, one has to implement the ``skeleton" with an oriented boundary
so that, mainly the $U(N)$ ribbons have typically the same structure
as in the previous situation. The graphs $T_i$'s and $F_i$'s, according to complex vertices $V^c_{\sigma,i}$,
have two different orientations.
We will denote the complex tadpoles and four point 1PI graphs
by $T^c_{i,s}$ and $F^c_{i,s}$, the meaning of $i$ remaining the same as
above, but the index $s=\pm$ is fixed according to the orientation 
(sign $+$ is affected to the counterclockwise rotation in the ``bubble",
see for instance Figures \ref{fig5} and \ref{fig6}).
We find the combinatorial factors 
(including sum over colour indices) and their amplitude sums,
with $S(m):= \sum_{p\in \mathbb{N}^2} G_{mp}$,
\begin{eqnarray}
&&K^c_{T_{1,+}}=2N=K^{c}_{T_{1,-}},\;\;\;
K^{c}_{T_{2,+}}= K^{c}_{T_{2,-}}=K^{c}_{T_{3,+}}=  K^{c}_{T_{3,-}}=2;\\
&& 
S_{T_{1,+}}(m)= -\frac{\lambda_2}{2 }S(m), \;\;\;
 S_{T_{1,-}}(n)= -\frac{\lambda_1}{2 }S(n), \;\;\;
   S_{T_{2,+}}(m)= -\frac{\lambda_1}{2 }S(m),\\
&&S_{T_{2,-}}(n)=  -\frac{\lambda_2}{2 }S(n),\;\;\;
S_{T_{3,+}}(m)= -\frac{\lambda_\times}{2 }S(m),\;\;\;
S_{T_{3,-}}(n)= -\frac{\lambda_\times}{2 }S(n);\\
&&{\cal K}^{c}_{F_{1,+}}=2^3 N={\cal K}^{c}_{F_{1,-}},
\;\;
{\cal K}^{c}_{F_{2,+}}=2^3= {\cal K}^{c}_{F_{2,-}},\;\;
{\cal K}^{c}_{F_{3,+}}=2^3={\cal K}^{c}_{F_{3,-}},\\
&&{\cal K}^{c}_{F_{4,+}}=2^3={\cal K}^{c}_{F_{4,-}},\;\;
{\cal K}^{c}_{F_{5,+}}=2^3={\cal K}^{c}_{F_{5,-}},\;\;
{\cal K}^{c}_{F_{6,+}}=2^3={\cal K}^{c}_{F_{6,-}};\\
&&
{\cal S}_{F_{1,+}}= \tilde\lambda_2 \;{\cal S}(m,k),\;\;
{\cal S}_{F_{1,-}}=\tilde\lambda_1 \;{\cal S}(n,l),\;\;
{\cal S}_{F_{2,+}}=\tilde\lambda_1 \;{\cal S}(m,k),\\
&&{\cal S}^{c}_{F_{2,-}}=\tilde\lambda_2 \;{\cal S}(n,l),\;\;
{\cal S}_{F_{3,+}}=\tilde\lambda_{\times}\, {\cal S}(m,k),\;\;
{\cal S}_{F_{3,-}}=\tilde\lambda_{\times}\, {\cal S}(n,l),\\
&&{\cal S}_{F_{4,+}}=\tilde\lambda_{12}\,{\cal S}(m,k),\;\;
{\cal S}_{F_{4,-}}=\tilde\lambda_{12}\,{\cal S}(n,l),\;\;
{\cal S}_{F_{5,+}}=\tilde\lambda_{2\times}\,{\cal S}(m,k),\\
&&{\cal S}_{F_{5,-}}=\tilde\lambda_{1\times}\,{\cal S}(n,l),\;\;
{\cal S}_{F_{6,+}}=\tilde\lambda_{1\times}\,{\cal S}(m,k),\;\;
{\cal S}_{F_{6,-}}= \tilde\lambda_{2\times}\,{\cal S}(n,l).
\end{eqnarray}
The below expressions are deduced by direct calculation at one loop
\begin{eqnarray}
&&Z^c= 1-(\lambda_1 + \lambda_2 N+\lambda_\times)S^{(1)},\\
&&\Gamma^c_{4,||,1}(0,0,0,0)= - \lambda_1 + 
(\lambda^2_2(1+N) + 2 \lambda_1 \lambda_2 +2  \lambda_2 \lambda_\times+ \lambda_\times^2)S^{(1)},\\
&&\Gamma^c_{4,||,2}(0,0,0,0)= - \lambda_2 +
(\lambda^2_1(1+N) + 2 \lambda_1 \lambda_2 +2  \lambda_1 \lambda_\times+ \lambda_\times^2)S^{(1)},\\
&&\Gamma^c_{4,\times}(0,0,0,0)= - \lambda_\times +2(\lambda_1+\lambda_2)\lambda_\times S^{(1)},
\end{eqnarray}
so that renormalized coupling flows can be inferred
\begin{eqnarray}
-\lambda^c_{r,||,1}  &=& - \lambda_1 - (2 \lambda^2_1 + 2 \lambda_1 \lambda_2 (N-1)+2 \lambda_1 \lambda_\times\\
&&
-\lambda^2_2(1+N)-2  \lambda_2 \lambda_\times-\lambda_\times^2)S^{(1)},\\
-\lambda^c_{r,||,2}&=& - \lambda_2 - (2 \lambda^2_2 \,N  +2 \lambda_2 \lambda_\times
-\lambda^2_1(1+N)- 2  \lambda_1 \lambda_\times - \lambda_\times^2)S^{(1)},\\
-\lambda^c_{\times} &=&-\lambda_\times -( 2 \lambda_\times^2-2 \lambda_2 \lambda_\times(1-N)).
\end{eqnarray}
Keeping $\lambda_{\times}$ fixed to zero, $ \lambda_1= \lambda_2 $ and for $N>1$,
we get $\lambda^c_{r,1}> \lambda_{bare,1} $ and  
$\lambda^c_{r,2} < \lambda_{bare,2} $  and, consequently,
the ``parallel" vertex $V_1$ determines an UV asymptotic freedom while 
$V_2$ takes the opposite direction of the triviality issue.  If we impose the equations 
$\lambda^c_{r,i}= \lambda_{bare,i} $, $i=1,2,\times$, we find the solutions
\begin{eqnarray}
[\lambda_{\times}=0,\;\; N=1,\;\; \lambda_{1}=\pm \lambda_2],
\quad 
[ N=2,\;\; \lambda_{1}=\lambda_2=-\lambda_{\times}].
\end{eqnarray}
The first set of parameters hints the complex renormalizable GWm \cite{diser2} 
only if $\lambda_1=+\lambda_2$. The second set of parameters 
implies an unstable model in the ``crossed" vertex sector if we assume that $\lambda_{1}>0$,
but it  is actually a new color model without Lg and possessing
a fixed point in the RG flow of all of its renormalized coupling constants.
This solution can be seen as the complex counterpart of the previous solution 
for $N=2$ found in the real situation.

\section{Conclusion}
\label{Sect4}

This letter has considered and analyzed the $O(N)$ and $U(N)$ GWm 
and computed their $\beta$-functions at one loop of perturbation. 
The real and complex NC color models are characterized by two and three kinds of vertices, respectively.
Their RG flows have been determined at one loop in order to find associated 
the $\beta$-functions. Real and complex GWm's are encoded as a limit $N=1$.
This study  has revealed the UV asymptotic freedom of a particular class of models with $N>1$
and only ``parallel" couplings. For $N=0$, the triviality problem or Lg of the ordinary $\phi^4_4$ has shown up again.

\section*{Acknowledgments}

J. B. G. thanks  the ANR Program ``GENOPHY" and the  Daniel Iagolnitzer Foundation (France)
for a research grant in the LPT-Orsay, Paris Sud XI.  The authors thanks Razvan Gurau (LPT-Orsay, Paris-Sud XI) 
and Jacques Magnen (Centre de Physique th\'eorique, Ecole Polytechnique, Palaiseau)
for very helpful discussions.

\end{document}